\documentclass[aps,prd,reprint,superscriptaddress,showpacs,showkeys]{revtex4-1}

\usepackage{graphicx}
\usepackage{natbib}

\usepackage{enumitem}
\setenumerate{itemindent=2em,leftmargin=0pt,itemsep=0pt,partopsep=0pt}
\setitemize[1]{itemindent=2em,leftmargin=0pt,itemsep=2pt,partopsep=0pt,topsep=0pt}

\begin{document}


\title{Limits on light WIMPs with a germanium detector at 177 eVee threshold at the China Jinping Underground Laboratory}


\affiliation{Key Laboratory of Particle and Radiation Imaging (Ministry of Education) and Department of Engineering Physics, Tsinghua University, Beijing 100084}
\affiliation{College of Physical Science and Technology, Sichuan University, Chengdu 610064}
\affiliation{Department of Nuclear Physics, China Institute of Atomic Energy, Beijing 102413}
\affiliation{School of Physics, Nankai University, Tianjin 300071}
\affiliation{NUCTECH Company, Beijing 100084}
\affiliation{YaLong River Hydropower Development Company, Chengdu 610051}
\affiliation{Institute of Physics, Academia Sinica, Taipei 11529}
\affiliation{Department of Physics, Banaras Hindu University, Varanasi 221005}

\author{S.K. Liu}
\affiliation{College of Physical Science and Technology, Sichuan University, Chengdu 610064}
\affiliation{Key Laboratory of Particle and Radiation Imaging (Ministry of Education) and Department of Engineering Physics, Tsinghua University, Beijing 100084}
\author{Q. Yue}
\email{Corresponding author: yueq@mail.tsinghua.edu.cn}
\affiliation{Key Laboratory of Particle and Radiation Imaging (Ministry of Education) and Department of Engineering Physics, Tsinghua University, Beijing 100084}
\author{K.J. Kang}
\affiliation{Key Laboratory of Particle and Radiation Imaging (Ministry of Education) and Department of Engineering Physics, Tsinghua University, Beijing 100084}
\author{J.P. Cheng}
\affiliation{Key Laboratory of Particle and Radiation Imaging (Ministry of Education) and Department of Engineering Physics, Tsinghua University, Beijing 100084}
\author{H.T. Wong}
\altaffiliation{Participating as a member of TEXONO Collaboration}
\affiliation{Institute of Physics, Academia Sinica, Taipei 11529}
\author{Y.J. Li}
\affiliation{Key Laboratory of Particle and Radiation Imaging (Ministry of Education) and Department of Engineering Physics, Tsinghua University, Beijing 100084}
\author{S.T. Lin}
\affiliation{College of Physical Science and Technology, Sichuan University, Chengdu 610064}
\affiliation{Institute of Physics, Academia Sinica, Taipei 11529}
\author{J.P. Chang}
\affiliation{NUCTECH Company, Beijing 100084}
\author{N.~Chen}
\affiliation{Key Laboratory of Particle and Radiation Imaging (Ministry of Education) and Department of Engineering Physics, Tsinghua University, Beijing 100084}
\author{Q.H.~Chen}
\affiliation{Key Laboratory of Particle and Radiation Imaging (Ministry of Education) and Department of Engineering Physics, Tsinghua University, Beijing 100084}
\author{Y.H. Chen}
\affiliation{YaLong River Hydropower Development Company, Chengdu 610051}
\author{Y.C. Chuang}
\altaffiliation{Participating as a member of TEXONO Collaboration}
\affiliation{Institute of Physics, Academia Sinica, Taipei 11529}
\author{Z. Deng}
\affiliation{Key Laboratory of Particle and Radiation Imaging (Ministry of Education) and Department of Engineering Physics, Tsinghua University, Beijing 100084}
\author{Q. Du}
\affiliation{College of Physical Science and Technology, Sichuan University, Chengdu 610064}
\author{H. Gong}
\affiliation{Key Laboratory of Particle and Radiation Imaging (Ministry of Education) and Department of Engineering Physics, Tsinghua University, Beijing 100084}
\author{X.Q. Hao}
\affiliation{Key Laboratory of Particle and Radiation Imaging (Ministry of Education) and Department of Engineering Physics, Tsinghua University, Beijing 100084}
\author{H.J. He}
\affiliation{Key Laboratory of Particle and Radiation Imaging (Ministry of Education) and Department of Engineering Physics, Tsinghua University, Beijing 100084}
\author{Q.J.~He}
\affiliation{Key Laboratory of Particle and Radiation Imaging (Ministry of Education) and Department of Engineering Physics, Tsinghua University, Beijing 100084}
\author{H.X.~Huang}
\affiliation{Department of Nuclear Physics, China Institute of Atomic Energy, Beijing 102413}
\author{T.R. Huang}
\altaffiliation{Participating as a member of TEXONO Collaboration}
\affiliation{Institute of Physics, Academia Sinica, Taipei 11529}
\author{H. Jiang}
\affiliation{Key Laboratory of Particle and Radiation Imaging (Ministry of Education) and Department of Engineering Physics, Tsinghua University, Beijing 100084}
\author{H.B. Li}
\altaffiliation{Participating as a member of TEXONO Collaboration}
\affiliation{Institute of Physics, Academia Sinica, Taipei 11529}
\author{J.M. Li}
\affiliation{Key Laboratory of Particle and Radiation Imaging (Ministry of Education) and Department of Engineering Physics, Tsinghua University, Beijing 100084}
\author{J. Li}
\affiliation{Key Laboratory of Particle and Radiation Imaging (Ministry of Education) and Department of Engineering Physics, Tsinghua University, Beijing 100084}
\author{J. Li}
\affiliation{NUCTECH Company, Beijing 100084}
\author{X. Li}
\affiliation{Department of Nuclear Physics, China Institute of Atomic Energy, Beijing 102413}
\author{X.Q. Li}
\affiliation{School of Physics, Nankai University, Tianjin 300071}
\author{X.Y.~Li}
\affiliation{School of Physics, Nankai University, Tianjin 300071}
\author{Y.L.~Li}
\affiliation{Key Laboratory of Particle and Radiation Imaging (Ministry of Education) and Department of Engineering Physics, Tsinghua University, Beijing 100084}
\author{H.Y. Liao}
\altaffiliation{Participating as a member of TEXONO Collaboration}
\affiliation{Institute of Physics, Academia Sinica, Taipei 11529}
\author{F.K. Lin}
\altaffiliation{Participating as a member of TEXONO Collaboration}
\affiliation{Institute of Physics, Academia Sinica, Taipei 11529}
\author{L.C. L\"{u}}
\affiliation{Key Laboratory of Particle and Radiation Imaging (Ministry of Education) and Department of Engineering Physics, Tsinghua University, Beijing 100084}
\author{H. Ma}
\affiliation{Key Laboratory of Particle and Radiation Imaging (Ministry of Education) and Department of Engineering Physics, Tsinghua University, Beijing 100084}
\author{S.J. Mao}
\affiliation{NUCTECH Company, Beijing 100084}
\author{J.Q. Qin}
\affiliation{Key Laboratory of Particle and Radiation Imaging (Ministry of Education) and Department of Engineering Physics, Tsinghua University, Beijing 100084}
\author{J. Ren}
\affiliation{Department of Nuclear Physics, China Institute of Atomic Energy, Beijing 102413}
\author{J. Ren}
\affiliation{Key Laboratory of Particle and Radiation Imaging (Ministry of Education) and Department of Engineering Physics, Tsinghua University, Beijing 100084}
\author{X.C.~Ruan}
\affiliation{Department of Nuclear Physics, China Institute of Atomic Energy, Beijing 102413}
\author{M.B.~Shen}
\affiliation{YaLong River Hydropower Development Company, Chengdu 610051}
\author{L.~Singh}
\altaffiliation{Participating as a member of TEXONO Collaboration}
\affiliation{Institute of Physics, Academia Sinica, Taipei 11529}
\affiliation{Department of Physics, Banaras Hindu University, Varanasi 221005}
\author{M.K. Singh}
\altaffiliation{Participating as a member of TEXONO Collaboration}
\affiliation{Institute of Physics, Academia Sinica, Taipei 11529}
\affiliation{Department of Physics, Banaras Hindu University, Varanasi 221005}
\author{A.K. Soma}
\altaffiliation{Participating as a member of TEXONO Collaboration}
\affiliation{Institute of Physics, Academia Sinica, Taipei 11529}
\affiliation{Department of Physics, Banaras Hindu University, Varanasi 221005}
\author{J. Su}
\affiliation{Key Laboratory of Particle and Radiation Imaging (Ministry of Education) and Department of Engineering Physics, Tsinghua University, Beijing 100084}
\author{C.J. Tang}
\affiliation{College of Physical Science and Technology, Sichuan University, Chengdu 610064}
\author{C.H. Tseng}
\altaffiliation{Participating as a member of TEXONO Collaboration}
\affiliation{Institute of Physics, Academia Sinica, Taipei 11529}
\author{J.M.~Wang}
\affiliation{YaLong River Hydropower Development Company, Chengdu 610051}
\author{L. Wang}
\affiliation{Key Laboratory of Particle and Radiation Imaging (Ministry of Education) and Department of Engineering Physics, Tsinghua University, Beijing 100084}
\affiliation{College of Physical Science and Technology, Sichuan University, Chengdu 610064}
\author{Q.~Wang}
\affiliation{Key Laboratory of Particle and Radiation Imaging (Ministry of Education) and Department of Engineering Physics, Tsinghua University, Beijing 100084}
\author{S.Y. Wu}
\affiliation{YaLong River Hydropower Development Company, Chengdu 610051}
\author{Y.C. Wu}
\affiliation{Key Laboratory of Particle and Radiation Imaging (Ministry of Education) and Department of Engineering Physics, Tsinghua University, Beijing 100084}
\author{Y.C. Wu}
\affiliation{NUCTECH Company, Beijing 100084}
\author{Z.Z. Xianyu}
\affiliation{Key Laboratory of Particle and Radiation Imaging (Ministry of Education) and Department of Engineering Physics, Tsinghua University, Beijing 100084}
\author{R.Q. Xiao}
\affiliation{Key Laboratory of Particle and Radiation Imaging (Ministry of Education) and Department of Engineering Physics, Tsinghua University, Beijing 100084}
\author{H.Y. Xing}
\affiliation{College of Physical Science and Technology, Sichuan University, Chengdu 610064}
\author{F.Z. Xu}
\affiliation{Key Laboratory of Particle and Radiation Imaging (Ministry of Education) and Department of Engineering Physics, Tsinghua University, Beijing 100084}
\author{Y. Xu}
\affiliation{School of Physics, Nankai University, Tianjin 300071}
\author{X.J. Xu}
\affiliation{Key Laboratory of Particle and Radiation Imaging (Ministry of Education) and Department of Engineering Physics, Tsinghua University, Beijing 100084}
\author{T.~Xue}
\affiliation{Key Laboratory of Particle and Radiation Imaging (Ministry of Education) and Department of Engineering Physics, Tsinghua University, Beijing 100084}
\author{C.W. Yang}
\affiliation{College of Physical Science and Technology, Sichuan University, Chengdu 610064}
\author{L.T. Yang}
\affiliation{Key Laboratory of Particle and Radiation Imaging (Ministry of Education) and Department of Engineering Physics, Tsinghua University, Beijing 100084}
\author{S.W. Yang}
\altaffiliation{Participating as a member of TEXONO Collaboration}
\affiliation{Institute of Physics, Academia Sinica, Taipei 11529}
\author{N. Yi}
\affiliation{Key Laboratory of Particle and Radiation Imaging (Ministry of Education) and Department of Engineering Physics, Tsinghua University, Beijing 100084}
\author{C.X. Yu}
\affiliation{School of Physics, Nankai University, Tianjin 300071}
\author{H. Yu}
\affiliation{Key Laboratory of Particle and Radiation Imaging (Ministry of Education) and Department of Engineering Physics, Tsinghua University, Beijing 100084}
\author{X.Z. Yu}
\affiliation{College of Physical Science and Technology, Sichuan University, Chengdu 610064}
\author{X.H. Zeng}
\affiliation{YaLong River Hydropower Development Company, Chengdu 610051}
\author{Z. Zeng}
\affiliation{Key Laboratory of Particle and Radiation Imaging (Ministry of Education) and Department of Engineering Physics, Tsinghua University, Beijing 100084}
\author{L.~Zhang}
\affiliation{NUCTECH Company, Beijing 100084}
\author{Y.H. Zhang}
\affiliation{YaLong River Hydropower Development Company, Chengdu 610051}
\author{M.G. Zhao}
\affiliation{School of Physics, Nankai University, Tianjin 300071}
\author{W. Zhao}
\affiliation{Key Laboratory of Particle and Radiation Imaging (Ministry of Education) and Department of Engineering Physics, Tsinghua University, Beijing 100084}
\author{Z.Y. Zhou}
\affiliation{Department of Nuclear Physics, China Institute of Atomic Energy, Beijing 102413}
\author{J.J. Zhu}
\affiliation{College of Physical Science and Technology, Sichuan University, Chengdu 610064}
\author{W.B. Zhu}
\affiliation{NUCTECH Company, Beijing 100084}
\author{X.Z. Zhu}
\affiliation{Key Laboratory of Particle and Radiation Imaging (Ministry of Education) and Department of Engineering Physics, Tsinghua University, Beijing 100084}
\author{Z.H. Zhu}
\affiliation{YaLong River Hydropower Development Company, Chengdu 610051}

\collaboration{CDEX Collaboration}
\noaffiliation



\date{\today}

\begin{abstract}
  The China Dark Matter Experiment reports results on light WIMP dark matter searches at the China Jinping Underground Laboratory with a germanium detector array with a total mass of 20~g. The physics threshold achieved is 177 eVee  (``ee" represents electron equivalent energy) at 50\% signal efficiency. With 0.784 kg-days of data, exclusion region on spin-independent coupling with the nucleon is derived, improving over our earlier bounds at WIMP mass less than 4.6 GeV.
\end{abstract}


\maketitle

\section{Introduction}

Compelling evidence from astroparticle physics and cosmology indicates that dark matter constitutes about 27\% of the energy density of our Universe \cite{Beringer2012,*PlanckCollaboration2013a}. Weakly interacting massive particles (WIMPs, denoted by $\chi$) are the leading candidate for cold dark matter \cite{Kelso2012}. It is expected that WIMPs would interact with normal matter through elastic scattering. Direct detection of WIMPs has been attempted with different detector technologies \cite{Lewin1996}. 
The anomalous excess of unidentified events at low energy with the DAMA \cite{DAMACollaboration2011,*Bernabei2010}, CoGeNT \cite{Aalseth2011,*Aalseth2013,*CoGeNT_AM_2014}, CRESST-II \cite{Angloher2012} and CDMS (Si) \cite{CDMS_Si_PRL_2013} data has been interpreted as signatures of light WIMPs. They are however inconsistent with the null results from XENON \cite{Aprile2012a}, TEXONO \cite{TEXONO_2013,*TEXONO_BS_2014}, CDMSlite \cite{CDMSLite_2014_PRL}, LUX \cite{LUX_2013_PRL}, SuperCDMS \cite{SuperCDMS_2014_PRL} and CDEX-1 \cite{CDEX_1kg_2013,*CDEX_1kg_hardware,*CDEX_1kg_2014_arxiv} experiments.
It is crucial to continue probing WIMPs with lower mass achievable by available techniques.

Our earlier measurements \cite{CDEX_1kg_2013,*CDEX_1kg_hardware,*CDEX_1kg_2014_arxiv} have provided the first results on low-mass WIMPs from the China Dark Matter Experiment phase I (CDEX-1). With a 994 g point-contact germanium detector, an energy threshold of 400 eVee was achieved. 
The experiment was performed at the China Jinping Underground Laboratory (CJPL) \cite{CDEX_introduction}, which was inaugurated at the end of 2010.
With a rock overburden of more than 2400 m giving rise to a measured muon flux of 61.7 y$^{-1}\cdot$m$^{-2}$ \cite{WuYC2013}, CJPL provides an ideal location for low-background experiments.

We report final results of the ``CDEX-0" experiment at CJPL, which is based on a pilot measurement with an existing prototype Ge detector with sub-keV energy threshold at a few gram modular mass. The experimental setup, candidate event selection procedures and constraints on WIMP-nucleon spin-independent elastic scattering are discussed in the subsequent sections.

\begin{figure}[h]
  \includegraphics[width=1.0\linewidth]{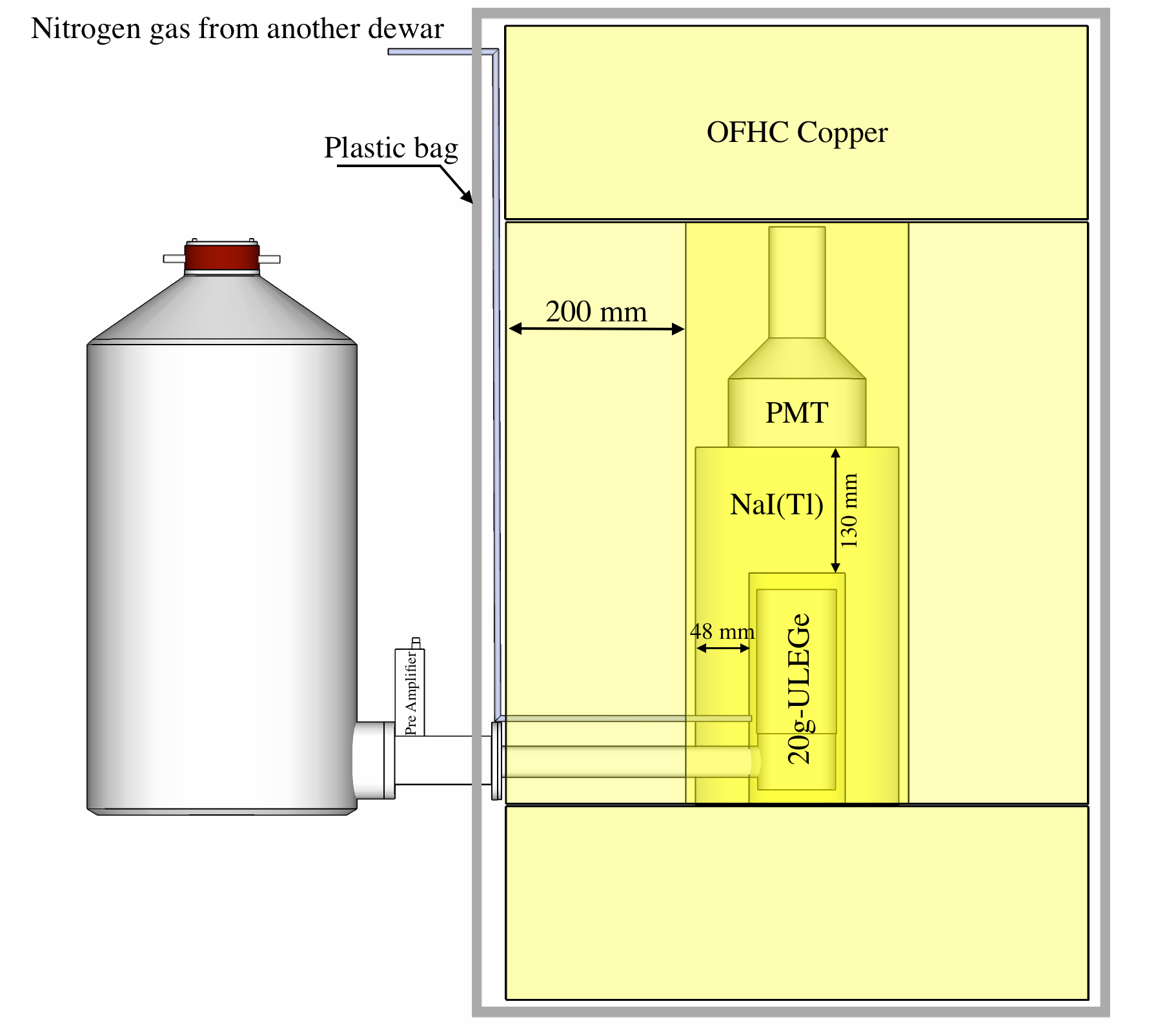}
  \caption{\label{fig:facility} Schematic diagram of the experimental setup which includes the germanium detector array and NaI(Tl) anti-Compton detector, as well as the enclosing OFHC Cu shielding. The entire structure is placed inside a passive shielding system described in Ref.\cite{CDEX_introduction}.}
\end{figure}

\begin{figure*}[htb]
  \includegraphics[width=1.0\linewidth]{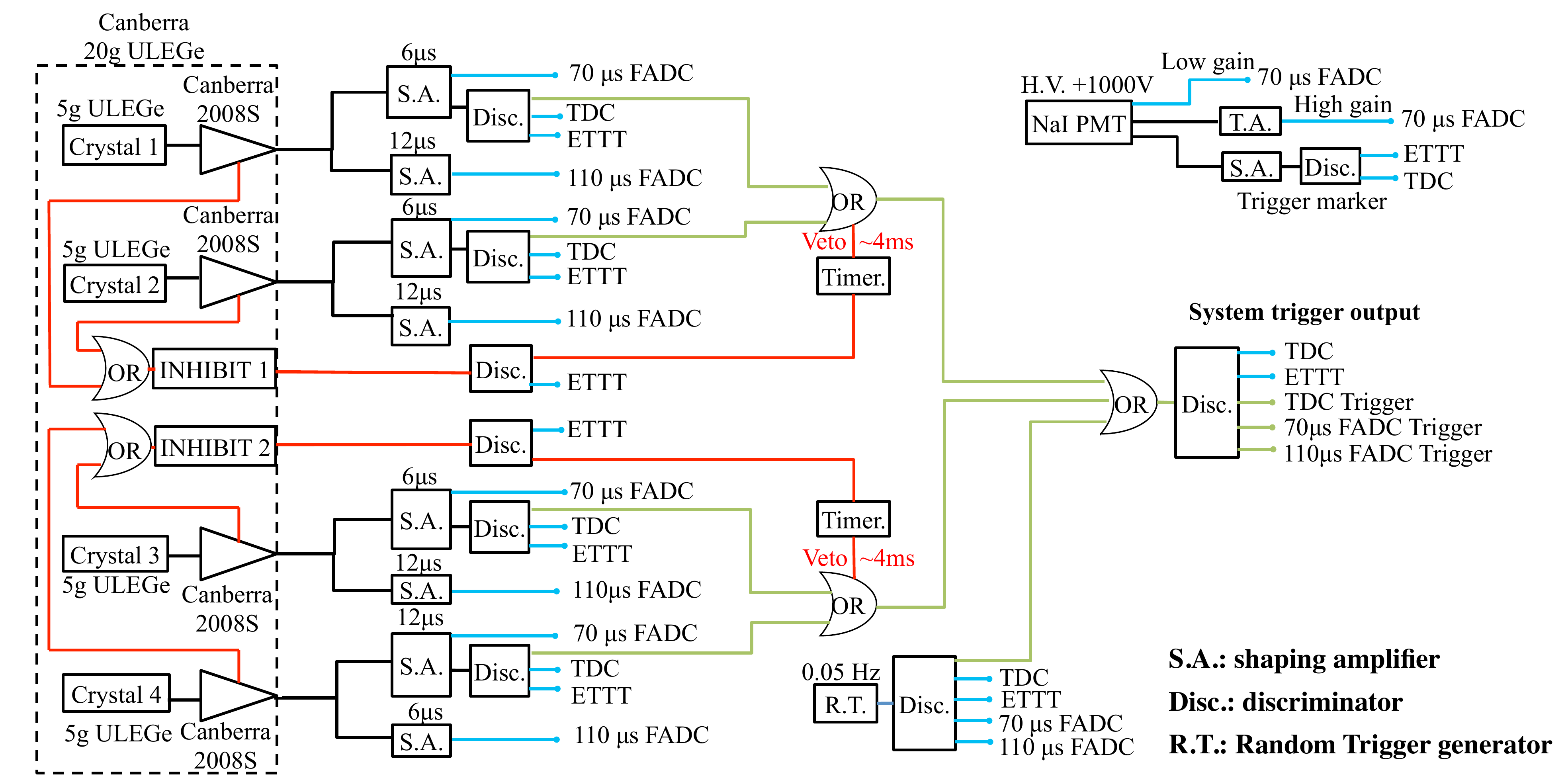}%
  \caption{\label{fig:daq} Schematic diagram of the electronics and the DAQ system of the germanium array and the NaI(Tl) detector.}
\end{figure*}

\begin{figure}[htb]
  \includegraphics[width=1.0\linewidth]{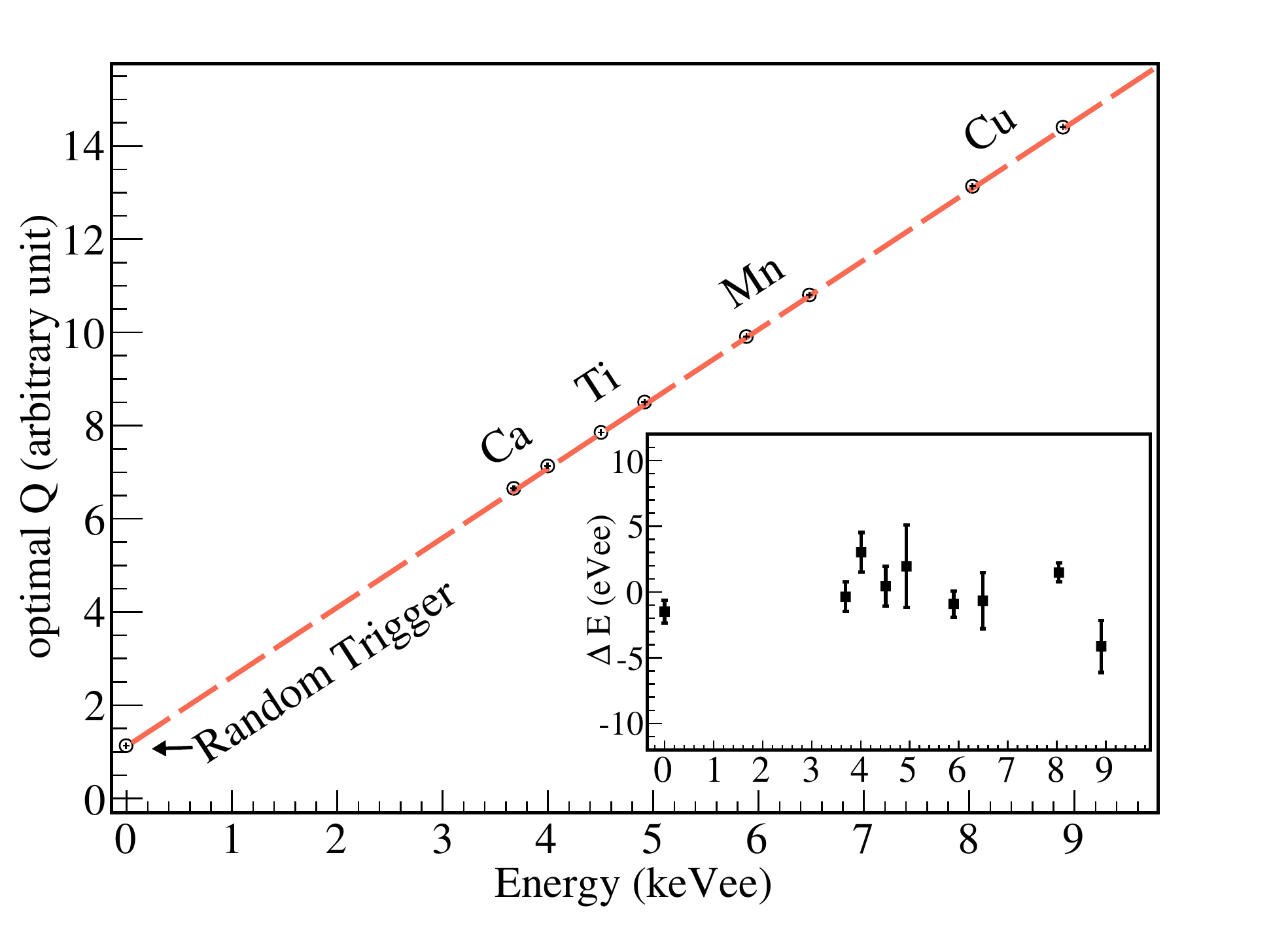}
  \caption{\label{fig:calibrate}  Calibration line relating the optimal Q measurements from SA$_{6}$ with the known energies from X-ray sources. The error bars are smaller than the data point size. The energy difference between the energy derived from the calibration and the real energy are depicted in the inset.}
\end{figure}

\begin{figure}[htbp]
  \includegraphics[width=1.0\linewidth]{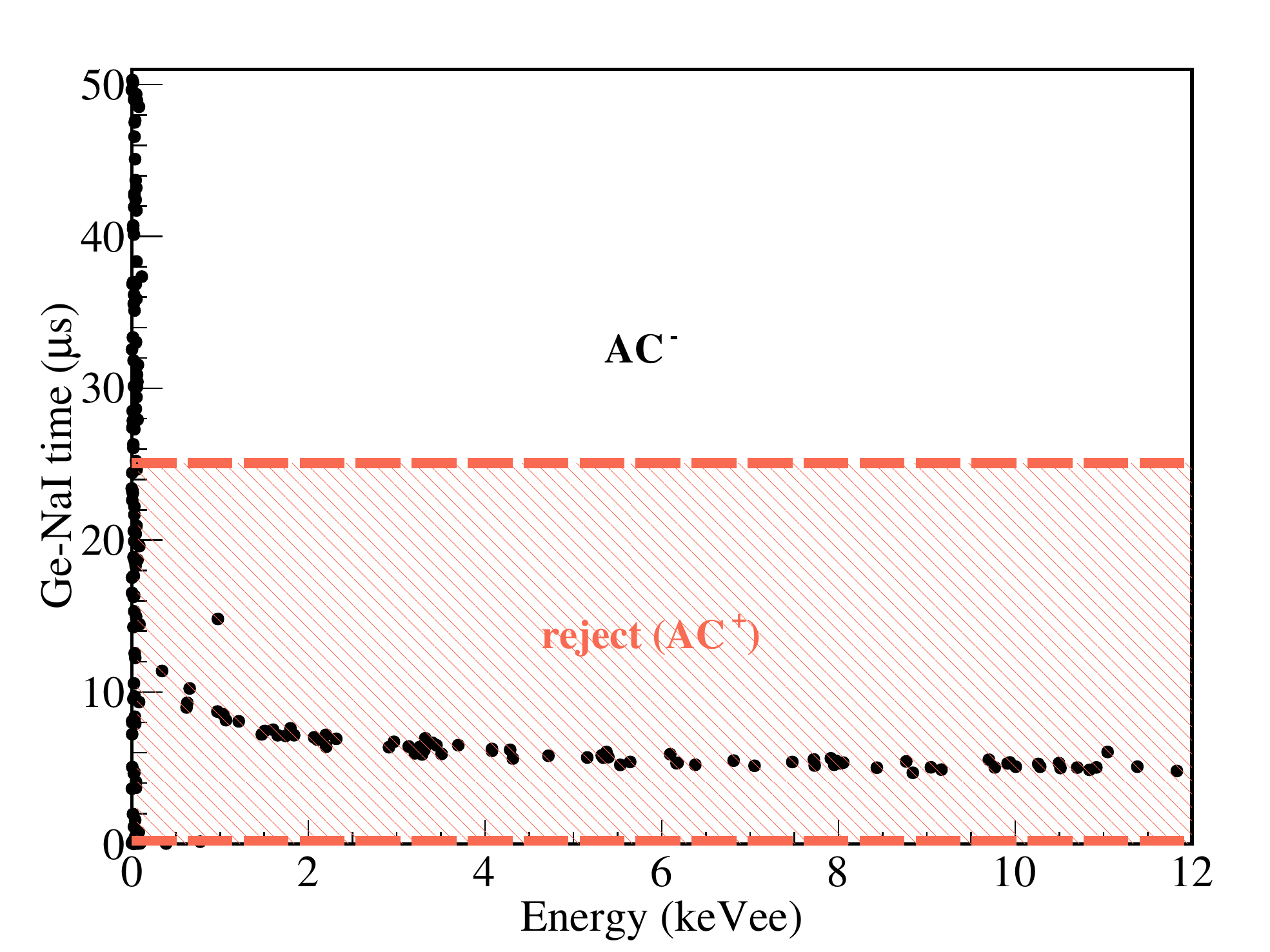}
  \caption{\label{fig:AC_cut}  Scatter plots of the difference between Ge and NaI(Tl) timing versus measured energy along with AC$^{-}$ selection and rejected parameter space.}
\end{figure}

\begin{figure}[htb]
  \includegraphics[width=1.0\linewidth]{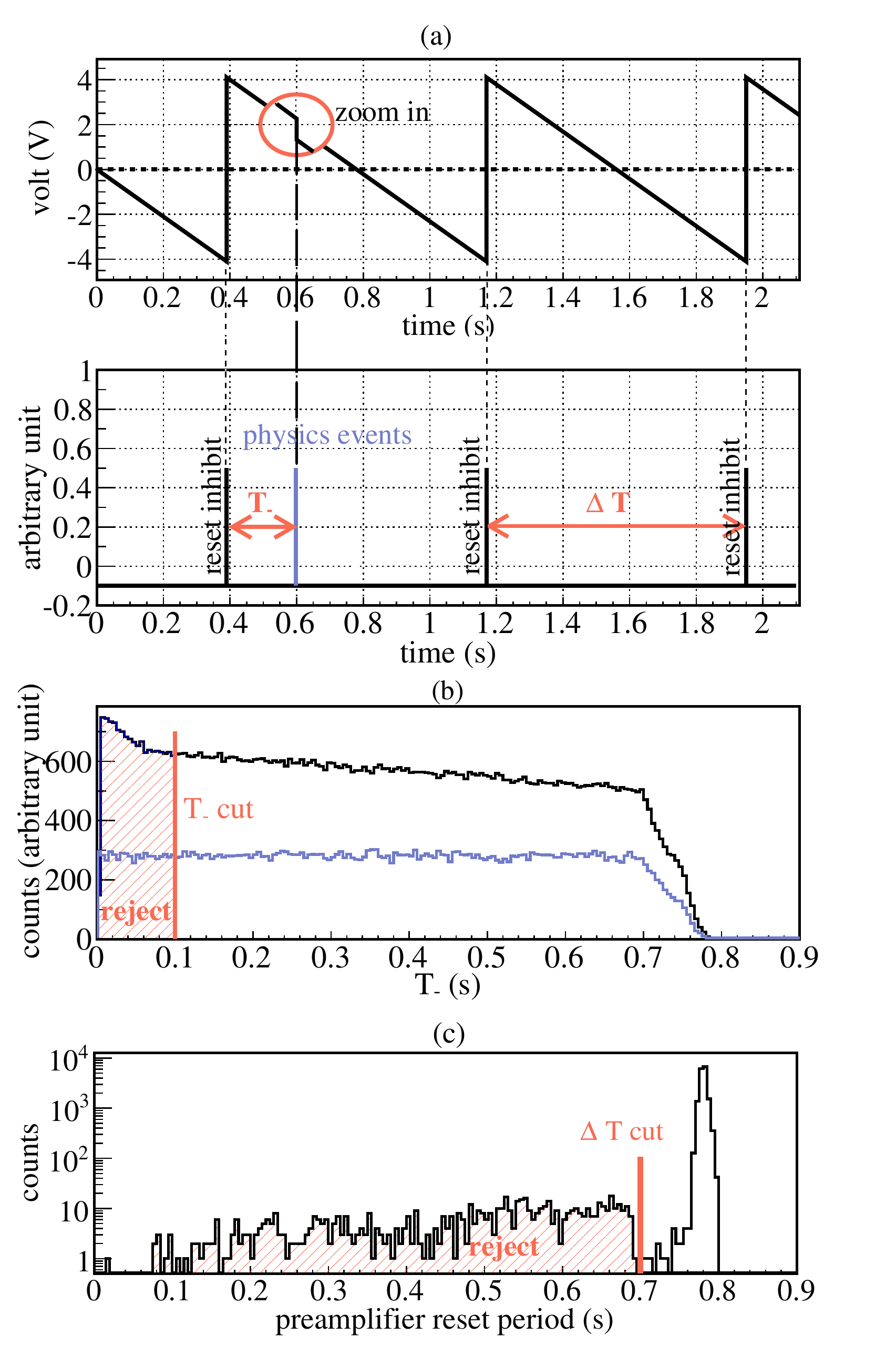}
  \caption{\label{fig:time_cuts}  (a) Raw signal from the reset preamplifier along with the timing of reset inhibit and a typical physics event. (b) The distributions of T$_{\text{-}}$ for random trigger events (blue) and background events (black)  are shown as well as the rejected parameter space. (c) The reset period cut and its rejected parameter space are displayed. }
\end{figure}

\begin{figure}[htb]
  \includegraphics[width=1.0\linewidth]{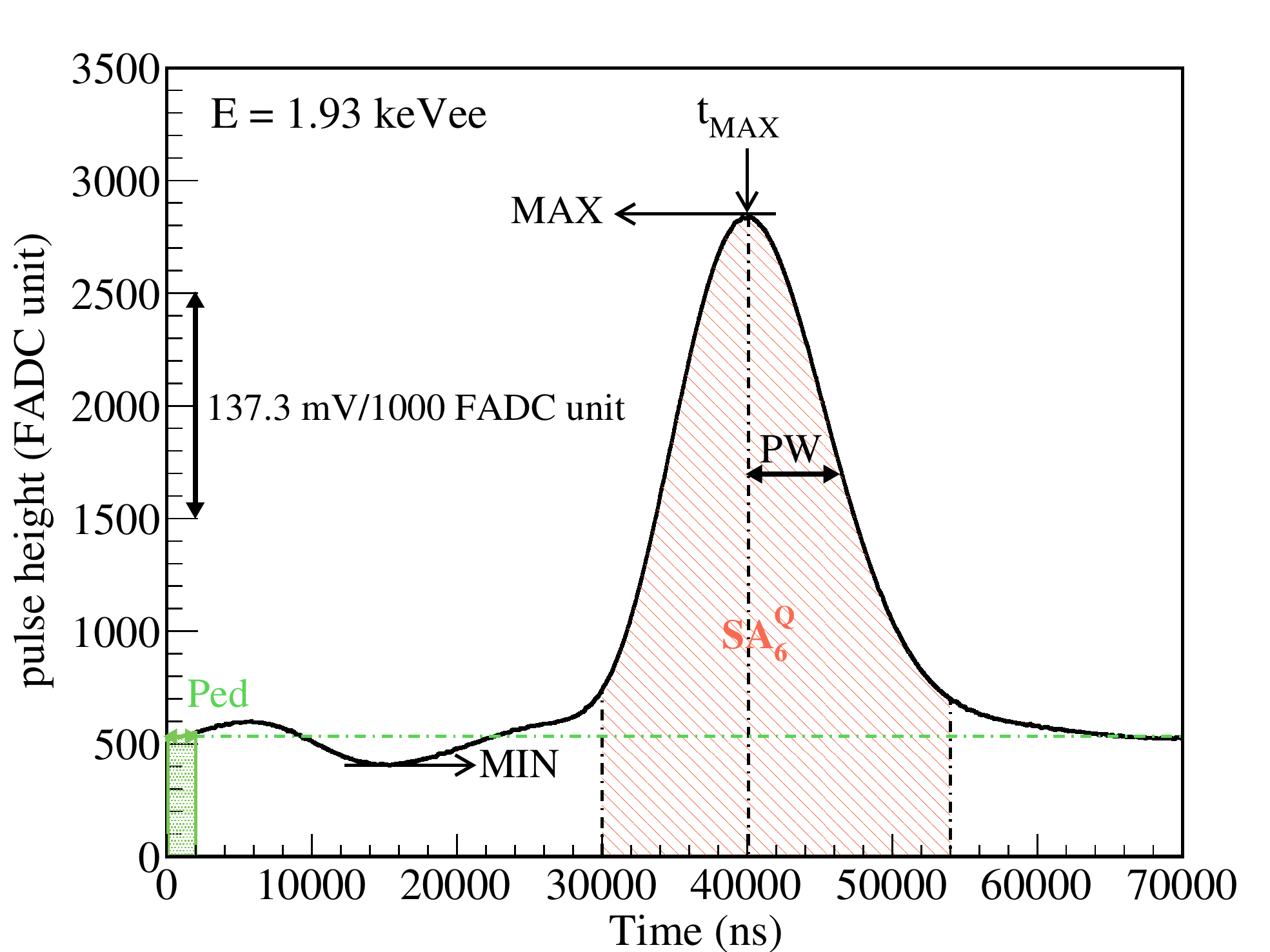}
  \caption{\label{fig:pulse_para}  Pulse shape parameters for PN Selection: Ped is the average of the first 200 time bins; MIN and MAX are the minima and maxima of the pulses, respectively, t$_{\text{MAX}}$ is the location of the maxima relative to the trigger instant and PW characterizes the pulse width. Energy is defined by the area SA$^{\text{Q}}_{6}$.}
\end{figure}

\begin{figure}[htbp]
  \includegraphics[width=1.0\linewidth]{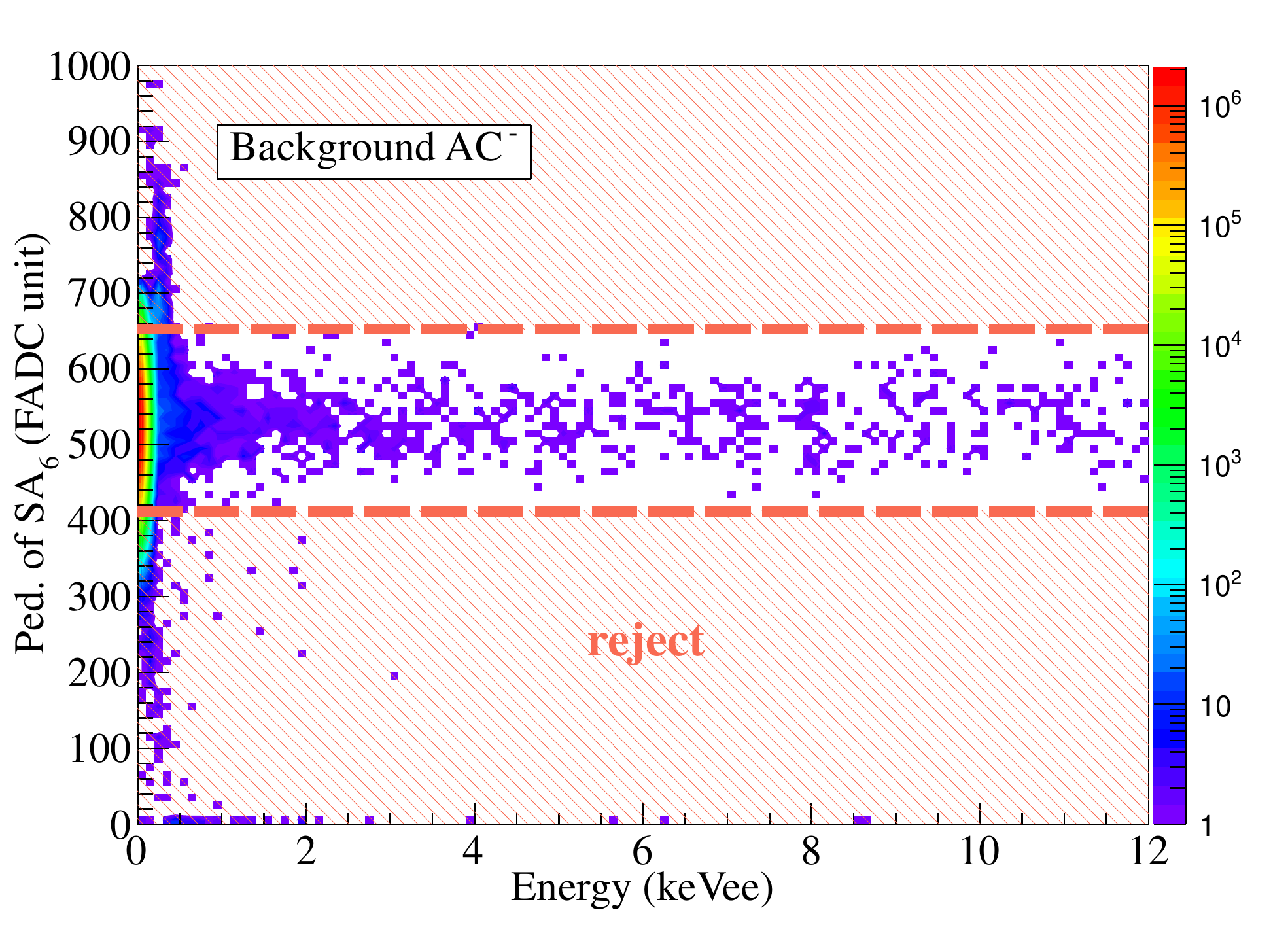}
  \caption{\label{fig:PNi_cut} Energy-independent PN$_{\text{i}}$ cut on Ped of SA$_{6}$.
  } 
\end{figure}

\begin{figure*}[htb]
  \includegraphics[width=1.0\linewidth]{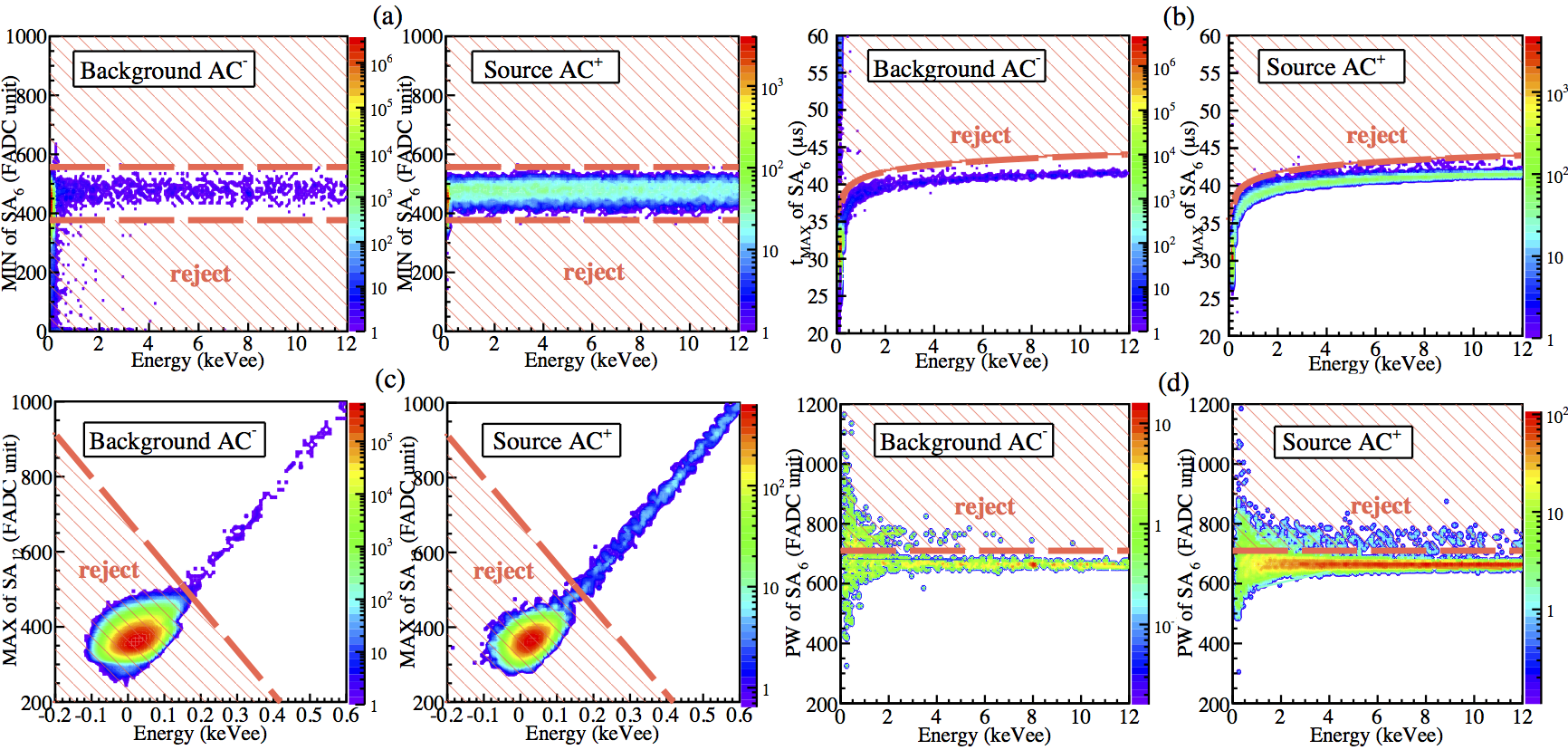}
  \caption{\label{fig:PNd_cuts} The energy-dependent PN$_{\text{d}}$ cuts: (a) MIN cut, (b)  t$_{\text{MAX}}$ cut, (c) MAX cut and (d) PW cut, based on the parameters defined in Fig.~\ref{fig:pulse_para}.}
\end{figure*}

\begin{figure}[htbp]
  \includegraphics[width=0.9\linewidth]{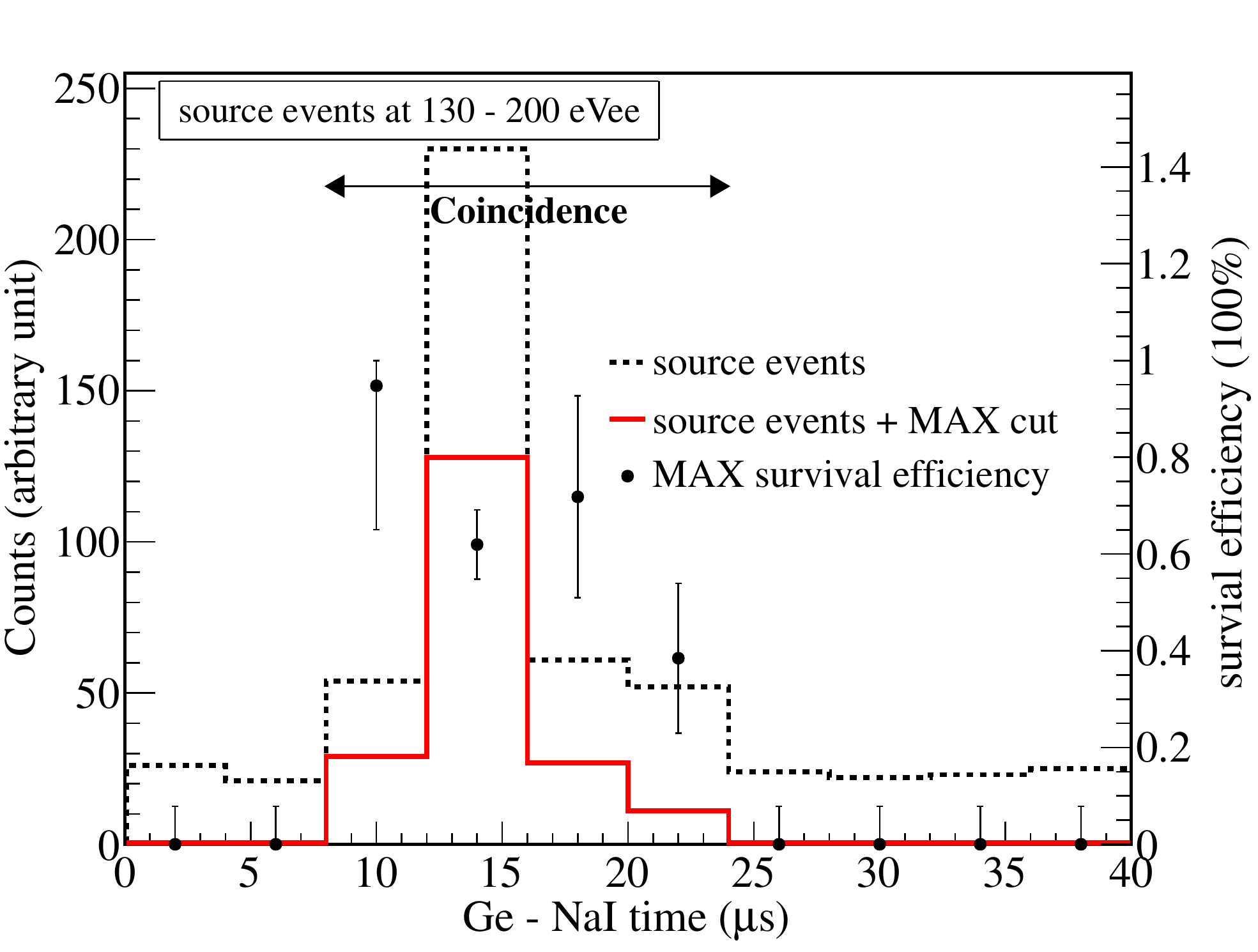}
  \caption{\label{fig:PN_work} The source events at 130-200 eVee versus the relative temporal distance between ULEGe triggers and AC signals are shown. The substantial value of survival efficiency at the coincidence time demonstrates the effectiveness of the MAX cut.}
\end{figure}

\begin{figure}[htbp]
  \includegraphics[width=1.0\linewidth]{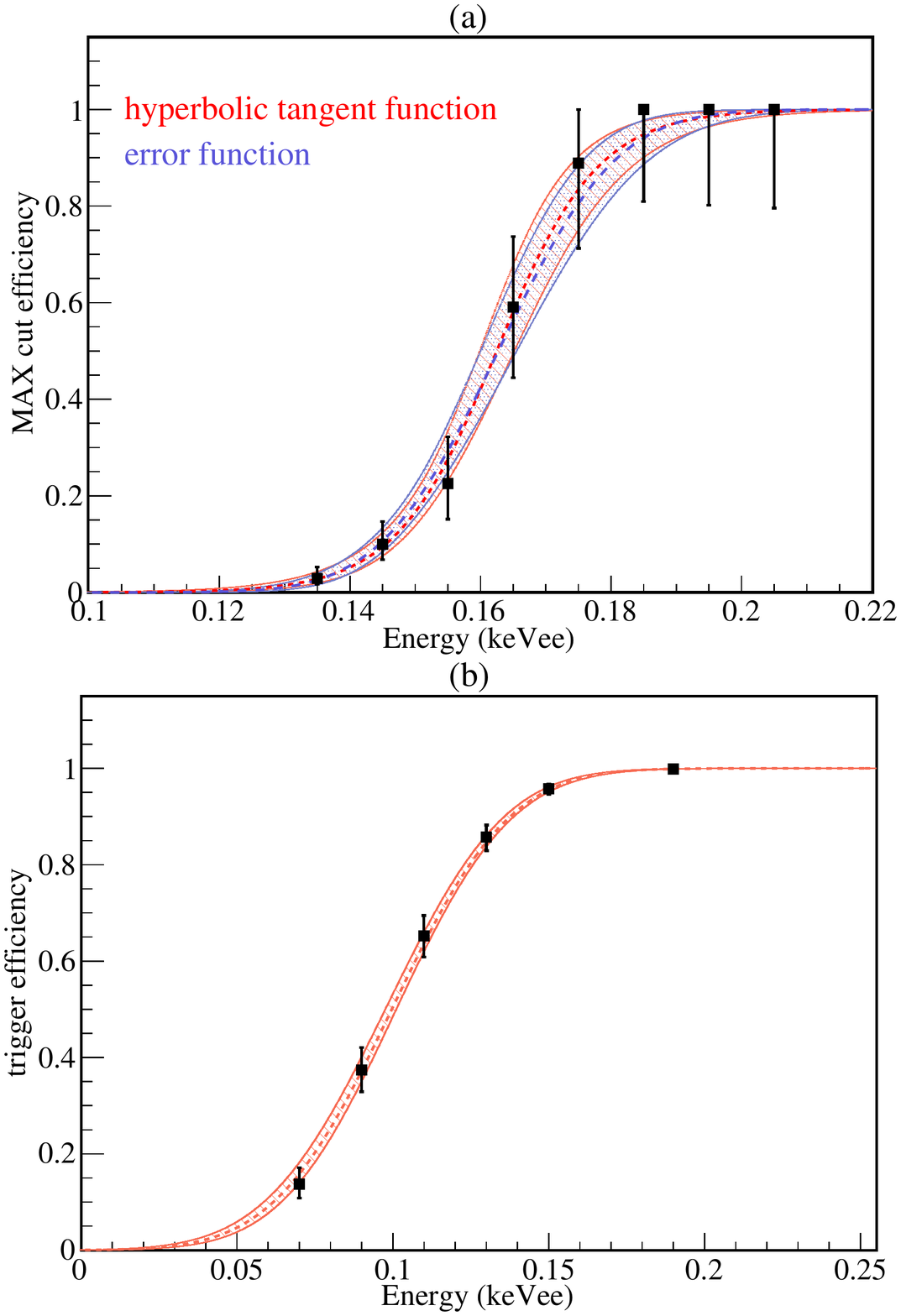}
  \caption{\label{fig:MAX_trig} (a) MAX cut efficiencies with two different fitting functions. (b) Trigger efficiencies derived from the source AC$^{+}$ events.}
\end{figure}

\begin{figure}[htb]
  \includegraphics[width=1.0\linewidth]{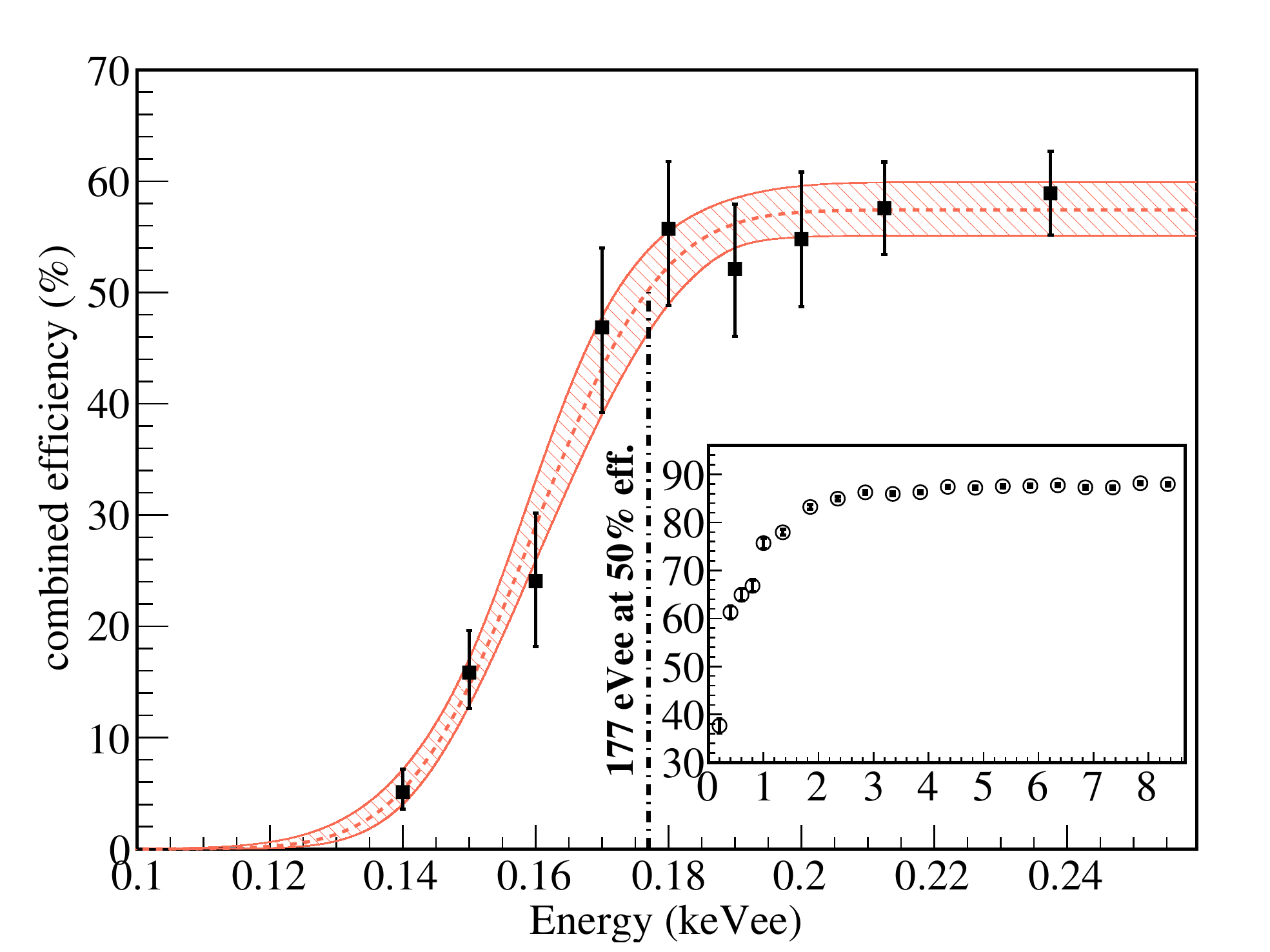}
  \caption{\label{fig:combined_eff} The combined efficiencies in the low energy range and in an extended energy range are depicted respectively. In the latter one the error bars are smaller than the data point size.}
\end{figure}

\begin{figure}[htbp]
  \includegraphics[width=1.0\linewidth]{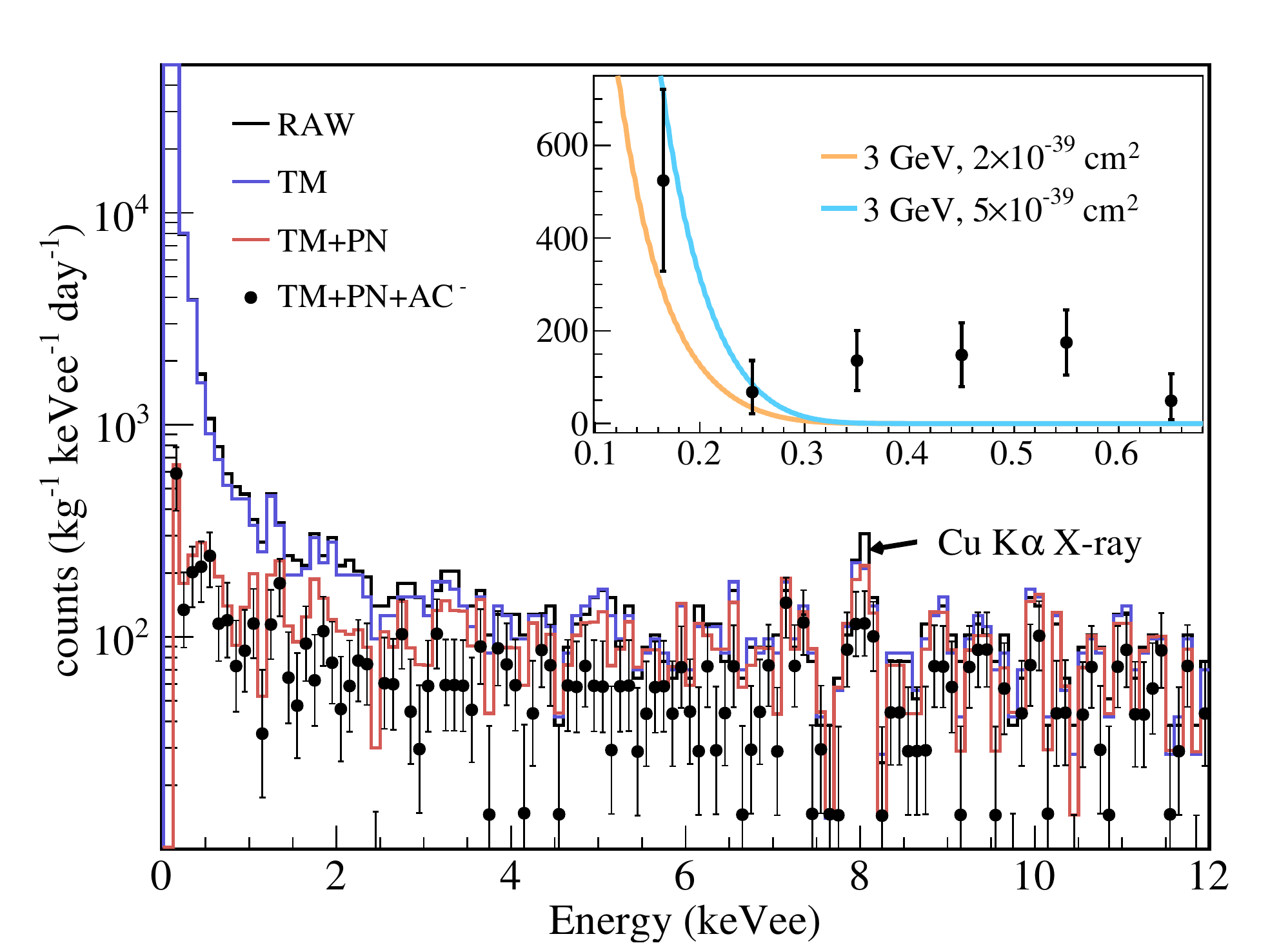}
  \caption{\label{fig:spec} Measured energy spectra of 20g-ULEGe, showing the raw spectra and those at the different stages of the analysis. The inset figure shows the low-energy spectrum after subtraction of a flat background due to high-energy $\gamma$ rays, superimposed with the predicted spectra for 3 GeV WIMPs with ${\sigma}^{\text{SI}}_{\chi{\text{N}}} = 2 \times 10 ^{-39}$ cm$^2$ and ${\sigma}^{\text{SI}}_{\chi{\text{N}}} = 5 \times 10 ^{-39}$ cm$^2$. }
\end{figure}

\section{Experimental Setup}

\begin{table}[htb]
\caption{\label{tab:crystal_performance} Measured parameters which characterize the resolution performance of the four crystals.}
\begin{ruledtabular}
\begin{tabular}{ccccc}
 Crystal&Pedestal&Noise&FWHM$^{a}$&Trigger efficiency\\
 number&rms&edge&of Ca x rays&at 50\%\\
 &(eVee)&(eVee)&at 3.69 keV (eVee)&(eVee)\\
  \hline
  1 & 51 & $\sim300$ & $169\pm3$ & $171\pm5$ \\ 
  2 & 33 & $\sim180$ & $123\pm3$ & $100\pm2$ \\
  3 & 64 & $\sim400$ & $186\pm6$ & $267\pm3$ \\  
  4 & 68 & $\sim250$ & $182\pm6$ & $146\pm4$\\
\end{tabular}
\leftline{$^{a}$Full width at half maximum (FWHM).}
\end{ruledtabular}
\end{table}

The CDEX-0 detector was previously used by the TEXONO experiment at a surface laboratory, where constraints on light WIMP were placed with a data set having a threshold of 220 eVee at 50\% signal efficiency \cite{TEXONO_2009}. The schematic design is depicted in Fig.~\ref{fig:facility}. 
The NaI(Tl) crystal scintillator whose threshold is about 10 keVee served as an anti-Compton (AC) 
 detector which enclosed the cryostat. The thickness of its side is 48 mm and top is 130 mm. The passive shielding system includes, from outside to inside, 1 m of polyethylene, 
20 cm of lead, 20 cm of borated polyethylene and 20 cm of OFHC (oxygen-free high-conductivity) copper. A plastic bag enclosing the OFHC copper was purged by nitrogen gas evaporated from the liquid nitrogen dewar.
The detailed information about the passive shielding system was described in Ref. \cite{CDEX_introduction}. 

The 20 g ultra low energy threshold germanium (ULEGe) detector consists of four n-type crystals. Every crystal with an active mass of 5 g has a semiplanar
configuration with a p$^+$ electrode on the outer surface, and a n$^+$ contact of a small diameter is used as the 
central electrode, from which signals are derived. 
The surface electrode is of $\mu$m thickness fabricated by boron-ion implantation.
The crystal array is encapsulated within the cryostat made of OFHC copper and the crystal center-to-center 
spacing is 14 mm. 
The cryostat end cap is made of carbon composite with the thickness of 0.6 mm allowing calibration with low-energy x rays outside.

The schematic diagram of the electronics and data acquisition (DAQ) system is shown in Fig.~\ref{fig:daq}. 
The n$^+$ contact signal is read out by low noise FET in the vicinity of the Ge crystal and fed into a reset preamplifier. 
Each crystal has its respective preamplifier and two identical outputs distributed to two 
shaping amplifiers at 6 $\mu$s (SA$_{\text{6}}$) and 12 $\mu$s (SA$_{\text{12}}$) shaping time. One output from 
SA$_{\text{6}}$ was fed into the discriminator to supply the 
 trigger for the DAQ system. 
 The signals were sampled and recorded by a 100 MHz flash analog-to-digital convertor (FADC). The recording time intervals were 70 $\mu$s and 110 $\mu$s for the signals at SA$_{\text{6}}$ and SA$_{\text{12}}$, respectively. 
 The photomultiplier tube (PMT) outputs from the AC detector at two different gain factors were also digitized. 
 A veto period of 4 ms is applied after every preamplifier reset to reject electronic-induced noise \cite{CDEX_1kg_2013}. 
 Events provided by a random trigger (RT) with a pulse generator at 0.05 Hz were also recorded for calibration and DAQ dead time measurements. These RT events are also used to derive the efficiencies of those analysis selection procedures which are uncorrelated with the pulse shape of the Ge signals.
  
The relative timing of the Ge and AC detectors were recorded by standard time to digital convertors (TDC) with 25 ps resolution and a full range of 52 $\mu$s.
In addition, the extended trigger time tag (ETTT) TDC extended the dynamic range to a full scale range up to 100 seconds maintaining the 25 ps resolution. It is used to detect long-duration temporal correlations, such as those between the Ge and the reset signal.

\begin{table}[htb]
\caption{\label{tab:all_efficiencies} Summary of candidate event selection procedures at two representative energy intervals. Listed are the individual and cumulative background survival fraction [$\lambda$(\%) and $\Pi \lambda$(\%), respectively] and the candidate signal efficiency [$\epsilon$(\%)], as well as ``Combined Efficiency" multiplying all efficiency factors together including trigger efficiency, TM, PN$_{\text{i}}$, PN$_{\text{d}}$ and AC$^{-}$ efficiencies.}
\begin{ruledtabular}
\begin{tabular}{lcc}
 Energy bin&130-200 eVee&600-700 eVee\\
  Raw background counts	& 5329 & 62  \\ 
  DAQ dead time (\%)		&\multicolumn{2}{c}{10.1$\pm0.2$}\\  
  \hline 
  Trigger efficiency (\%)		&96.6$\pm^{0.9}_{1.1}$ & 100 \\
  \hline
  \multicolumn{3}{c}{Timing Selection}\\
    	\multicolumn{1}{c}{$\lambda [ \Pi \lambda ]$(\%)}	&95.8 [95.8] & 79.0 [79.0] \\
	\multicolumn{1}{c}{$\epsilon$(\%)}	& \multicolumn{2}{c}{77.1$^{a}$$\pm0.2$}\\
  \hline
  \multicolumn{3}{c}{PN$_{\text{i}}$ cuts} \\
    	\multicolumn{1}{c}{$\lambda [ \Pi \lambda ]$(\%)}	&92.5 [88.6] & 73.5 [58.1] \\
	\multicolumn{1}{c}{$\epsilon$(\%)}	& \multicolumn{2}{c}{98.9$\pm0.2$} \\
  \hline
  \multicolumn{3}{c}{PN$_{\text{d}}$ cuts} \\
  MAX cut: & & \\
    	\multicolumn{1}{c}{$\lambda  [ \Pi \lambda ]$(\%)}	& 2.0 [1.8]& 100 [58.1]\\	
	\multicolumn{1}{c}{$\epsilon$(\%)}	& 53.9$\pm^{4.5}_{5.0}$  &100$\pm0.0$ \\
  MIN cut: & & \\
    	\multicolumn{1}{c}{$\lambda  [ \Pi \lambda ]$(\%)}	& 38.0 [0.7] & 69.4 [40.3] \\	
	\multicolumn{1}{c}{$\epsilon$(\%)}	& 95.4$\pm1.5$ & 99.8$\pm0.2$\\
  t$_{\text{MAX}}$ cut: & & \\
    	\multicolumn{1}{c}{$\lambda  [ \Pi \lambda ]$(\%)}	& 27.0 [0.5] & 50.0 [29.0] \\	
	\multicolumn{1}{c}{$\epsilon$(\%)}	&  96.3$\pm1.4$ & 93.1$\pm1.1$\\
  PW cut: & & \\
    	\multicolumn{1}{c}{$\lambda  [ \Pi \lambda ]$(\%)}	& 40.7 [0.2] & 55.6 [16.1] \\	
	\multicolumn{1}{c}{$\epsilon$(\%)}	& 71.7$\pm3.4$ & 79.2$\pm1.9$\\
  Combined PN$_{\text{d}}$ cuts: & & \\
    	\multicolumn{1}{c}{$\lambda  [ \Pi \lambda ]$(\%)}	&0.2 [0.2] & 27.8 [16.1]\\	
	\multicolumn{1}{c}{$\epsilon$(\%)}	& 35.5$\pm^{3.5}_{3.8}$ & 73.6$\pm2.0$\\
  \hline
  \multicolumn{3}{c}{Anti-Compton Selection} \\
    	\multicolumn{1}{c}{$\lambda  [ \Pi \lambda ]$(\%)}	&90.9 [0.2] & 60.0 [9.7]\\		\multicolumn{1}{c}{$\epsilon$(\%)}	& \multicolumn{2}{c}{100.0$\pm0.2$}\\
  \hline
  Combined Efficiency(\%) & 30.9$\pm^{3.0}_{3.3}$ & 66.3$\pm{1.8}$\\
  After-all-cuts counts & 10 & 6 \\
  After-all-cuts rate  & & \\	
  (kg$^{-1}$keV$^{-1}$day$^{-1}$) & \raisebox{1.0ex}[0pt]{590$\pm^{195}_{197}$} & \raisebox{1.0ex}[0pt]{115$\pm^{58}_{39}$}\\
\end{tabular}
\end{ruledtabular}
\leftline{$^{a}$ applied to a subset of 38.6\% of the data.}
\end{table}

At a total DAQ rate of 6.9 Hz, the DAQ live time was measured to be 89.9 \% by RT events. The anomalously large dead time was due to inefficient methods of hardware synchronization in the prototype DAQ system, which were fixed in our subsequent data taking. The optimal area from SA$_{\text{6}}$, which was defined as the light red shadow in Fig.6, was chosen as the energy measurements. Energy calibration was achieved by the external x-ray peaks from Ca, Mn, Ti, Cu which were produced by the x-ray generator illuminating a mixture of these elements, as displayed in Fig.~\ref{fig:calibrate} \cite{CDEX_20g_2013}.

 The zero energy was defined by the random trigger events. 
 The calibration uncertainties are 1.69 eVee and 1.70 eVee at 130 eVee and 1 keVee, respectively. The measured parameters which characterize the resolution of the four crystals are shown in Table~\ref{tab:crystal_performance}. Crystal $^{\#}$2 provided the best performance and was adopted for subsequent analysis. The data taking interval spans over 174 days from November 24, 2012 to September 18, 2013 with interruptions due to laboratory construction and hardware failure,  providing 0.784 kg-days of physics data.
Inclusion of data from the other crystals to the analysis would not provide substantial improvement to the final physics results.

Events in Ge crystal in coincidence (anticoincidence) with the AC detector are denoted as AC$^{+(-)}$, respectively. 
Physics events induced by $\gamma$-rays are selected by the AC$^{+}$ tag and are used to optimize the selection criteria and to evaluate the signal efficiencies. Two complementary data sets were adopted. A $^{60}$Co source placed external to the NaI-AC detector provided high statistics data giving accurate measurement of the efficiencies, while the \textit{in situ} low-background AC$^{+}$ data served as consistency cross-check.  

\section{Candidate Event Selection}

WIMP-induced interactions are characterized by being single-site events uncorrelated with other detector components, and having the same pulse shape as the events due to genuine physical processes. A series of data analysis criteria were adopted to select the $\chi$N events, and their corresponding signal efficiencies were measured. The details are discussed as follows, while the results are summarized in Table~\ref{tab:all_efficiencies}.

\begin{enumerate}
\item
Timing (TM) selection: The preamplifier reset induces noise events with definite timing structure. The timing profiles of the reset and the events are illustrated in Fig.~\ref{fig:time_cuts} (a). The timing distribution between an event and its previous reset, denoted by T$_{-}$, is depicted in Fig.~\ref{fig:time_cuts} (b) for physics events and RT. A cut of T$_{-} <$ 0.1 ms removes all reset-induced background. The reset period ($\Delta$T) distribution is shown in Fig.~\ref{fig:time_cuts} (c). The duration in which at least five consecutive periods are persistently below 0.7 s is  rejected. They correspond to temporary surge of leakage currents in the detector.  The TM selection is applied to a subset of 38.6\% of the data where the reset timing was recorded, in which the signal efficiency is derived from the survival  probability of the RT events to be 77.1 \%.

\item Anti-Compton (AC$^{-}$) selection: The time difference between the AC and the Ge-trigger instant is depicted in Fig.~\ref{fig:AC_cut}. The band corresponds to AC$^{+}$ events with coincidence of the Ge and NaI. The dependence with energy is due to the slow shaping pulse taking longer time at lower energy to cross a fixed threshold in the discriminator. The signal efficiency is 99.9 \% from the survival probability of the RT events.

\item Physics versus electronic noise (PN) selection: 
These events can be differentiated by their pulse shape parameters as defined in Fig.~\ref{fig:pulse_para}, where a 1.93 keVee particle pulse is shown. The energy-independent PN$_{\text{i}}$ selection is based on the pedestal (Ped) stability, as illustrated in Fig.~\ref{fig:PNi_cut}, where the signal efficiencies are derived by the survival of the RT events to be 98.9\%. The energy-dependent PN$_{\text{d}}$ cuts make use of the minima (MIN) and maxima (MAX) of the pulse, the location of the maxima (t$_{\text{MAX}}$) and the pulse width (PW), as depicted in Fig.~\ref{fig:PNd_cuts}. Events induced by physical processes have different distributions in these parameters from those of electronic noise. Physics events are defined by data taken with calibration $^{60}$Co $\gamma$ sources with AC$^{+}$ tag, also shown in Fig.~\ref{fig:PNd_cuts}. The selection parameter spaces are then applied to the candidate AC$^{-}$ events. The survival probabilities of the AC$^{+}$ samples provide measurements of selection efficiencies.
\end{enumerate}

While the MIN, t$_{\text{MAX}}$ and PW cuts would filter electronic noise and microphonics events, the MAX cut of Fig.~\ref{fig:PNd_cuts} (c) is the most important one to define the analysis threshold \cite{TEXONO_2009}. Physics and electronic noises events show different correlations between the area and amplitude of the pulse. The rejection criteria select physics events below the noise edge of 200 eVee, with efficiencies provided by the AC$^{+}$ samples. The timing distribution of Ge and NaI(Tl) for source events at 130-200 eVee energy is depicted in Fig.~\ref{fig:PN_work}, before and after the MAX cut. Events outside the coincidence region are rejected at 100\% efficiencies, showing the selection is indeed differentiating signal from electron noise events. Events in the coincidence range are kept, with the survival fractions representing the signal selection efficiencies. The variations with energy is depicted in Fig.~\ref{fig:MAX_trig} (a). Both hyperbolic tangent and error functions provide good and consistent descriptions to the data at the threshold region.

The trigger efficiencies are derived from the pulse shape of SA$_{6}$. The amplitude and its rms are derived from the source AC$^{+}$ events at energy above the noise edge. The zero-energy point is defined by RT events, and interpolations are made for energy in between, assuming Gaussian distribution. The trigger efficiency is the fraction of the amplitude distribution above the discriminator threshold \cite{TEXONO_2009}, as indicated in Fig.~\ref{fig:MAX_trig} (b). The 50\% trigger efficiency corresponds to 100$\pm$2 eVee.

The combined efficiencies of all cuts (including trigger efficiency, TM, PN$_{\text{i}}$, PN$_{\text{d}}$ and AC$^{-}$ ) are shown in Fig.~\ref{fig:combined_eff}. The physics  threshold is $177\pm5$ eVee corresponding to a combined efficiency at 50\%.
The signal efficiencies and background suppression factors at threshold and at a high-energy bin are summarized in Table II to illustrate and compare the effects of each process. 

\section{Constraints on Light WIMPs}

 The measured raw spectra and those at different stages of the analysis are depicted in Fig.~\ref{fig:spec}. Standard error propagation formulas are adopted, using the statistical uncertainties of the raw measurements as well as those on selection efficiencies from Fig.~\ref{fig:combined_eff}.
 The minimum energy is at 130 eVee, matching the first finite efficiency bin of Fig.~\ref{fig:combined_eff}.
 ~The only explicit structure observed was the 8.041 keV copper K$\alpha$ x ray.  They are produced from the interactions of high-energy $\gamma$ rays on the copper support structures in the vicinity of the Ge crystal, and hence a portion of these events are tagged by the AC$^{-}$ selection. The spectrum of events that survived all selection criteria is flat above 1.5 keVee, due to ambient radioactivity of high-energy $\gamma$ rays. The residual spectrum after subtraction of this background channel is depicted in the inset of Fig.~\ref{fig:spec}.

The energy spectra due to $\chi$N spin-independent interactions cannot be larger than the residual spectrum. The thickness of the surface inactive layer \cite{dead_layer_1,*dead_layer_2,*dead_layer_3} is only tens of micrometers and can be neglected. Upper limits on their cross sections (${\sigma}^{\text{SI}}_{\chi{\text{N}}}$) as a function of WIMP mass are derived, using the binned Poisson method \cite{Savage2009}. The input parameters include quenching factor provided by the TRIM program \cite{Ziegler2004,*TEXONO_2007}, coupled with a 10\% systematic error implied by the spread of the measured data at the recoil energy of 254 eV to 10 keV, standard WIMP halo assumption \cite{Donato1998}, conventional astrophysical models (local WIMP density of 0.3~$\text{GeV}\cdot\text{cm}^{-3}$ and Maxwellian velocity distribution with $\nu_0 = 220$ $\text{km}\cdot\text{s}^{-1}$, the escape velocity $\nu_{\text{esc}} = 544$~$\text{km}\cdot\text{s}^{-1}$) and energy resolution of detector derived from the calibration data.  

\begin{figure}[htb]
  \includegraphics[width=1.0\linewidth]{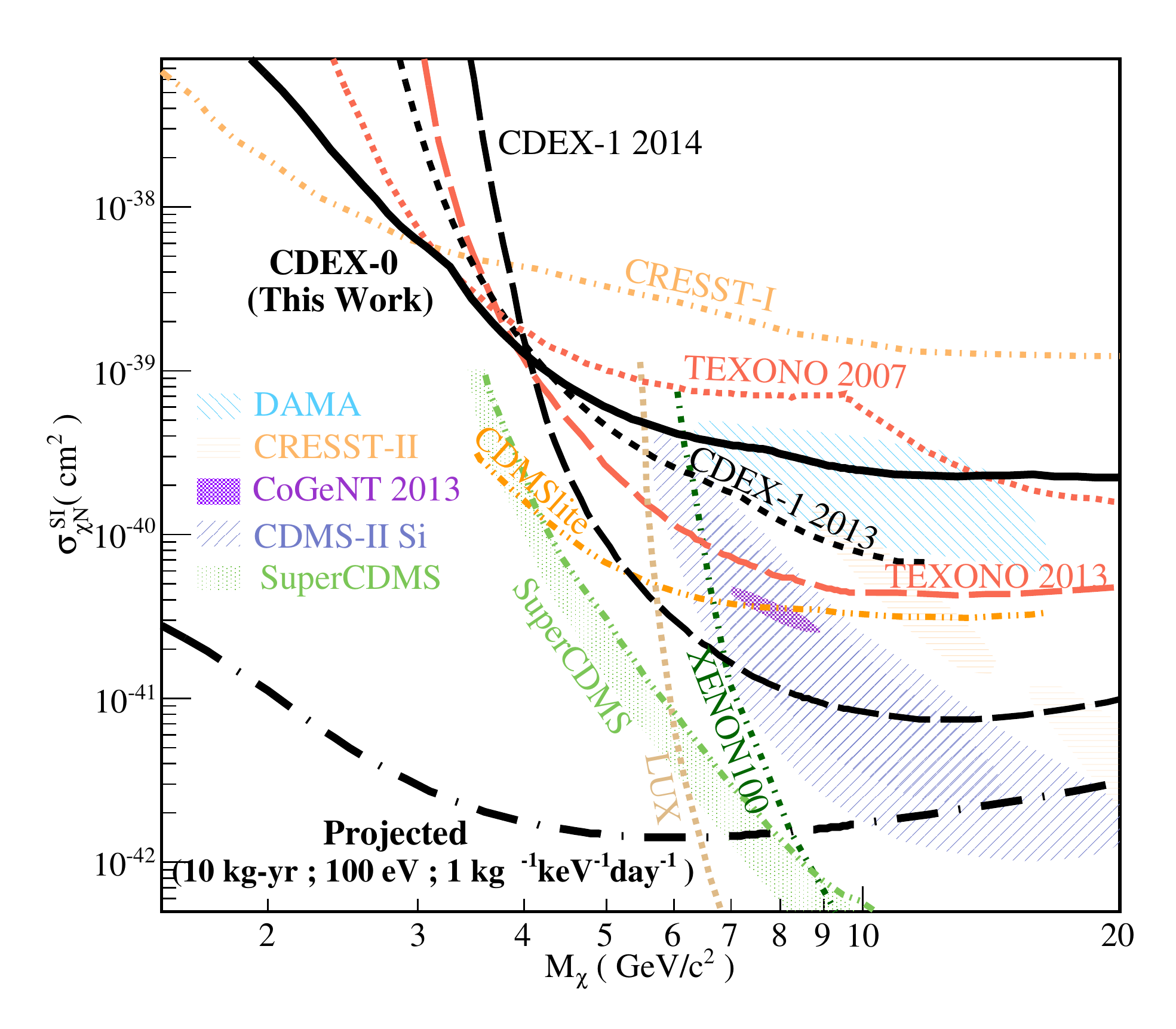}%
  \caption{\label{fig:ex-plot} Exclusion plot of spin-independent ${\chi}$N coupling at 90\% confidence level,
    superimposed with the results from other benchmark experiments. 
    Allowed regions given by CoGeNT \cite{Aalseth2011,*Aalseth2013,*CoGeNT_AM_2014}, DAMA/LIBRA \cite{DAMACollaboration2011,*Bernabei2010}, CDMS-II (Si) \cite{CDMS_Si_PRL_2013} and CRESST-II \cite{Angloher2012} are presented, as well as the exclusion limits from  XENON100 \cite{Aprile2012a}, TEXONO\cite{TEXONO_2009,TEXONO_2013,*TEXONO_BS_2014}, CDMSlite \cite{CDMSLite_2014_PRL}, LUX \cite{LUX_2013_PRL}, SuperCDMS \cite{SuperCDMS_2014_PRL}, CDEX-1 \cite{CDEX_1kg_2013,*CDEX_1kg_hardware,*CDEX_1kg_2014_arxiv} and CRESST-I \cite{Angloher2002}. The potential reach at indicated projected sensitivities with point-contact germanium detectors is also displayed.}
\end{figure}

The exclusion curve at 90\% confidence level is shown in Fig.~\ref{fig:ex-plot}, together with those of several selected experiments \cite{Aalseth2011,*Aalseth2013,DAMACollaboration2011,*Bernabei2010,CDMS_Si_PRL_2013,Angloher2012,Aprile2012a,CDMSLite_2014_PRL,SuperCDMS_2014_PRL,LUX_2013_PRL,TEXONO_2009,TEXONO_2013,*TEXONO_BS_2014,Angloher2002,CDEX_1kg_2013,*CDEX_1kg_hardware,*CDEX_1kg_2014_arxiv}.
The previous results of CRESST-I \cite{Angloher2002} and TEXONO \cite{TEXONO_2009} are reanalyzed using the currently-favored astrophysical parameters. Under this consistent analysis, this result improves over our earlier bounds from the same detector at a surface location \cite{TEXONO_2009} extending the low reach of light WIMPs to 2 GeV, and over the published limit \cite{CDMSLite_2014_PRL} at M$_{\chi}<$~3.5~GeV. 
The predicted $\chi$N recoil spectra due to the allowed (excluded) ${\sigma}^{\text{SI}}_{\chi{\text{N}}}$ at ${\text{m}}_{\chi}$ = 3 GeV are superimposed with the residual spectrum in the inset of Fig.~\ref{fig:spec}.


\section{Summary and Prospects}

The results presented in this article correspond to the first completed program of the pilot experiment at the new underground facility CJPL. Improved constraints are derived with a conventional Ge detector of good threshold response but only a few gram modular target mass. Novel p-type point-contact germanium detectors were developed in the past few years \cite{Luke1989,*P.S.Barbeau2007}, offering sub-keV energy threshold with kg-scale target such that the background level per unit mass is greatly reduced due to self-attenuation effects.  Dark matter experiments with this detector technique are being pursued at CJPL \cite{CDEX_1kg_2013,*CDEX_1kg_hardware,*CDEX_1kg_2014_arxiv} and elsewhere  \cite{Aalseth2011,*Aalseth2013,TEXONO_2013,*TEXONO_BS_2014}. The projected sensitivities of the realistic benchmark sensitivities of 100 eVee threshold at 1~kg$^{-1}$ keV$^{-1}$ day$^{-1}$ background level for 10 kg-year exposure is overlaid in Fig.~\ref{fig:ex-plot}. 




\begin{acknowledgments}
  This work was supported by the National Natural Science Foundation of China
  (Contracts No.10935005, No.~10945002, No. 11275107 and No. 11175099) and National Basic Research program of China (973 Program) (Contract No. 2010CB833006) and NSC 99-2112-M-001-017-MY3, and the Academia Sinica Principle Investigator Award 2011-2015 from Taiwan.
\end{acknowledgments}


\newpage

\bibliography{cdex0_20g_bib}

\end{document}